# Low-loss waveguides on Y-cut thin film lithium niobate: towards acousto-optic applications

LUTONG CAI[*], ASHRAF MAHMOUD AND GIANLUCA PIAZZA

*Department of Electrical and Computer Engineering, Carnegie Mellon University, Pittsburgh, PA, 15213, USA*

**Abstract:** We investigate the dependence of photonic waveguide propagation loss on the thickness of the buried oxide layer in Y-cut lithium niobate on insulator substrate to identify trade-offs between optical losses and electromechanical coupling of surface acoustic wave (SAW) devices for acousto-optic applications. Simulations show that a thicker oxide layer reduces the waveguide loss but lowers the electromechanical coupling coefficient of the SAW device. Optical racetrack resonators with different lengths were fabricated by argon plasma etching to experimentally extract waveguide losses. By increasing the thickness of the oxide layer from 1 μm to 2 μm, propagation loss of 2 μm (1 μm) wide waveguide was reduced from 1.85 dB/cm (3 dB/cm) to as low as 0.37 dB/cm (0.77 dB/cm), and, resonators with quality factor greater than 1 million were demonstrated. An oxide thickness of approximately 1.5 μm is sufficient to significantly reduce propagation loss due to leakage into the substrate and simultaneously attain good electromechanical coupling in acoustic devices. This work not only provides insights on the design and realization of low-loss photonic waveguides in lithium niobate, but most importantly offers experimental evidence on how the oxide thickness directly impacts losses and guides its selection for the synthesis of high-performance acousto-optic devices in Y-cut lithium niobate on insulator.

## 1. Introduction

Lithium niobite (LN), a versatile optical material which possesses outstanding properties including nonlinear optical, electro-optical, piezoelectric, and acousto-optical (AO) effects and a wide transparency range (350 nm ~ 5.2 μm) [1], has wide application in telecommunication networks, sensors, frequency conversions, quantum optics and microelectromechanical systems (MEMS) [2-5]. The emergence of lithium niobite on insulator (LNOI, thin film of LN bonded on low refractive index material) in the last two decades has led to rapid growth of LN photonics since its high index contrast waveguide structure enables dramatic reduction of the device footprint and enhancement of optical effects with respect to implementations in the bulk material [6]. Substantial cutting-edge research activities have flourished based on compact thin film LN photonic devices for a variety of applications [7-15]. For AO applications, LN features high elasto-optic coefficients like GaAs, but comes also with very strong piezoelectricity over many other materials (e.g., GaAs, ZnO, quartz…) [16], which results in acoustic devices that have a substantially higher electromechanical coupling coefficient [17]

However, compared to more mature material platforms like silicon on insulator (SOI) and other semiconductor materials, producing low loss waveguides in LNOI is extremely challenging. Dry etched waveguides have exhibited high propagation loss due to rough sidewalls or byproduct formed by chemical reaction [18,19]. Hybrid waveguides consisting of other material loaded on top of the LN thin film avoid etch issues but confine only a portion of the light in the active LN layer, hence they do not fully harness the material capabilities [20]. Only recently, low-loss subwavelength photonic waveguides have been attained by optimizing plasma etching conditions and coating over-cladding layers on X-cut and Z-cut LNOI [21,22]. Despite the significant progress, none of these prior demonstrations directly relates the achievable low loss in LNOI waveguides to the thickness of the buried oxide layer (BOX), but rather focus on the etch method to define the waveguides. It is important to note that, for AO devices, the buried oxide layer (BOX) has a direct impact on the electromechanical coupling coefficient ($k_t^2$) and thicker films of oxide negatively impact the performance of MHz devices such as those used for inertial sensing applications [13]. Therefore, it is well worth studying the dependence of the waveguide loss on the BOX thickness on LNOI as that would also impact the design of high-coupling AO devices. Differently from prior demonstrations, we study this dependence in Y-cut LN as it is one of the most appropriate cuts for the implementation of acousto-optic (AO) devices. In fact, in this cut, surface acoustic waves (SAW) can be efficiently excited by taking advantage of the high $k_t^2$ of the film and result in resonators with high quality factor [17].

Both simulations and experiments were conducted to study the dependence of waveguide losses on the thickness of the BOX layer. As predicted by simulations, the waveguide loss extracted from racetrack (RT) resonators decreased with increasing the BOX thickness. Low-loss waveguides (0.33 dB/cm) and high Q resonators (>1,000,000) were fabricated by argon plasma etching followed by RCA cleaning. Most importantly, the results show that a thickness of 1.5 μm could be used as a good compromise between photonic and acoustic performance.

## 2. Simulations

*2.1 Electromechanical coupling coefficient ($k_t^2$) as a function of oxide thickness*

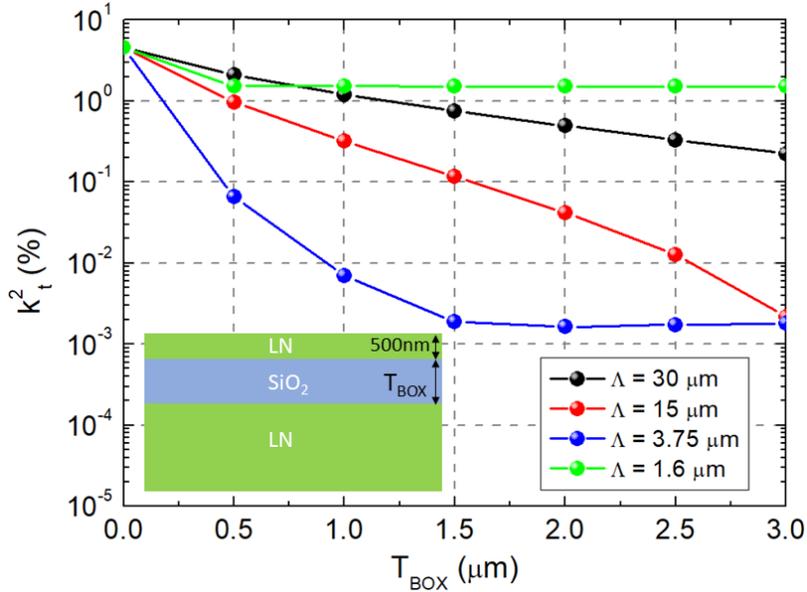

Fig. 1. Simulated electromechanical coupling coefficient ($k_t^2$) for various $T_{BOX}$ and acoustic wavelengths (Λ) on a YZ cut LNOI (see inset for wafer stack). The presence of oxide clearly has a deleterious impact on the device $k_t^2$. It is interesting to note how the impact changes with the specific wavelength. This is due to changes in penetration of the acoustic and electric fields in the thin films of LN, oxide and the thick LN substrate. At larger acoustic wavelengths, the oxide thickness is a small fraction of the acoustic devices. At intermediate wavelengths, the oxide becomes a dominant part of the active SAW and its impact on $k_t^2$ is more dramatic. At the smallest wavelength, the acoustic and electric fields are almost entirely confined in the thin film of LN and the oxide has a lower impact on $k_t^2$.

$k_t^2$, broadly defined as the ratio of the conversion between electrical and acoustic energy in a piezoelectric transducer, is an important figure of merit that determines the selection of a piezoelectric material for acoustic applications. $k_t^2$ is also of paramount importance in AO devices, since it ultimately determines the device size (inversely proportional to $k_t^2$) and the effectiveness with which an acoustic wave interacts with light. $k_t^2$ of SAW devices in bulk LN can be readily computed via numerical methods [17]. We extended the same methodology to derive the $k_t^2$ of SAW devices for LNOI by using finite element methods in COMSOL Multiphysics. We investigated the dependence of $k_t^2$ as a function of different BOX thicknesses ($T_{BOX}$) and acoustic wavelengths ($\Lambda$) for a Y-cut LNOI substrate. Y-cut LN was selected because it is one of the preferred cuts for SAW applications given the larger $k_t^2$ and lower acoustic losses that can be attained [17]. In this analysis, a z-propagating SAW wave was simulated in a thin film of LN having a thickness of 500 nm, a variable oxide thickness and a thick LN substrate (see Fig. 1 inset for material stack). The thickness of the top LN layer was set at 500 nm as the preferred value for the making of photonic waveguides with a well-confined quasi-TE$_{00}$ mode around 1550 nm. It is important to note that the presence of a thick LN substrate below the oxide is important for the effective excitation of large acoustic wavelength devices. In fact, if such substrate was not present or substituted with silicon, then large wavelength SAWs would not be effectively excited. The results of this analysis are shown in Fig. 1 where $k_t^2$ is plotted versus $T_{BOX}$ for different $\Lambda$. Although $T_{BOX}$ has a different impact on the $k_t^2$ depending on the specific wavelength (see Fig. 1 caption for further explanation), it is clear that the presence of $T_{BOX}$ has a deleterious effect on the electromechanical coupling of the SAW device. Therefore, depending on the specific application, the oxide thickness should be carefully selected and should not generally be made to exceed 2 µm except for particular acoustic wavelengths. In the following sections, we detail the impact of this same oxide thickness on photonic losses for the same Y-cut LNOI, so as to provide guidance in the design of high-performance AO devices.

## 2.2 Propagation losses as a function of oxide thickness

The schematic cross-section of a partially etched waveguide and the corresponding mode profile are shown in Fig. 2(a) and (b), respectively. The waveguide cross-section is the y-x crystal plane of the film and light propagates in the z-direction. We performed a full-vectorial finite difference simulation using a commercial software, Lumerical, to compute the propagation losses in this waveguide structure [23]. A perfectly matched layer (PML) boundary was used to enclose the waveguide region and absorb any incident electromagnetic field at its boundary. Both LN and BOX were defined as dielectric materials without intrinsic loss, hence the only accountable losses in the simulations were coming from the optical field decaying through the BOX and coupling to a radiation mode in the substrate (absorbed by the PML in the simulation framework). The contour plot in Fig. 2(c) shows that loss can be reduced by either increasing $T_{BOX}$ or the width of the rib (W). The rationale behind this behavior can be explained by looking at the optical field exponential decay in the substrate direction, $\sim \exp(-\alpha L)$, where $\alpha$ is the decay rate and L is the decay length. As $T_{BOX}$ increases, the decay length, L, increases, while as W increases a higher effective refractive index ($n_{eff}$) waveguide is formed thus enhancing the decay rate, $\alpha$. Losses of 0.01 dB/cm, 0.1 dB/cm, 1 dB/cm and 5 dB/cm are marked in the contour plot of Fig. 2(c) to roughly highlight what minimum oxide thickness is required in order to achieve such level of losses for a given waveguide width. We also calculated the dependence of losses on $T_{box}$ and $W$ for bent waveguides with a bending radius of 20 µm as shown in Fig. 2 (d). Clearly, because of the additional radiation losses the bent waveguides exhibit higher losses but similar trends. For large radius, e.g., the 100 µm we used in the devices reported in this work, the dependence of losses on $T_{box}$ and $W$ is practically the same as that for straight waveguides as illustrated in Fig. 2(c).

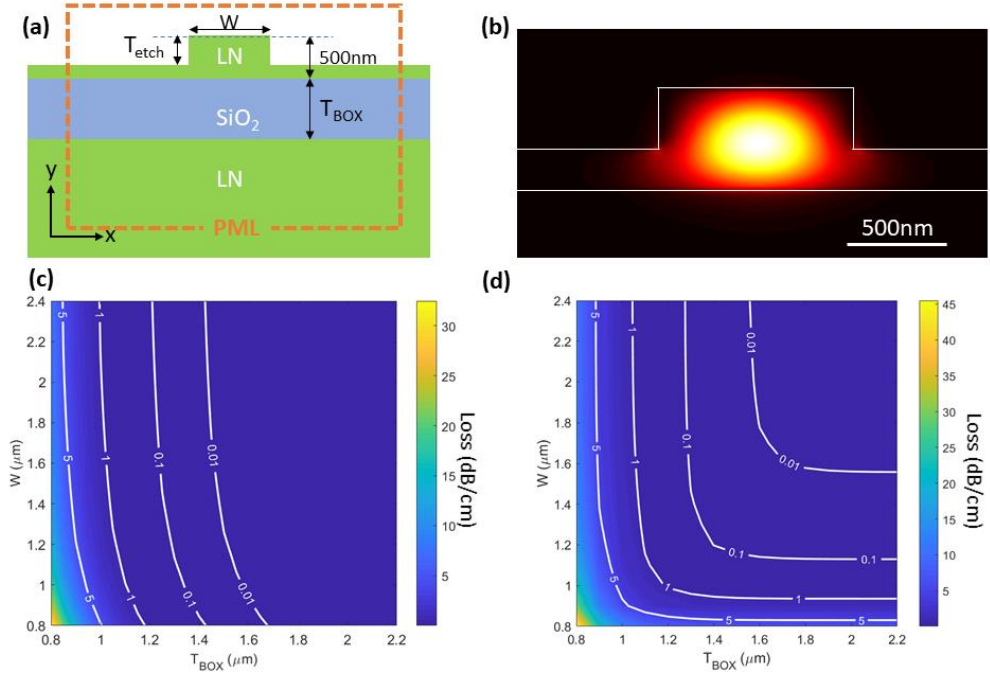

Fig. 2. (a) Schematic cross-section of the Y-cut LNOI rib waveguide. PML boundary condition was applied in the simulation to represent radiation losses into the substrate. (b) E intensity distribution of the fundamental TE-like mode of the waveguide with W = 1 μm and $T_{etch}$ = 300 nm. Calculated dependence of loss on W and $T_{BOX}$ for (c) straight waveguide and (d) bent waveguide (radius = 20 μm).

## 3. Experimental demonstration of low loss waveguides in Y-cut lithium niobate

Methods such as the cut-back and Fabry-Perot (FP) interference [24,25] are commonly used to extract waveguide losses. The cut-back method needs consistent coupling efficiency from the fiber to the waveguide to accurately extract loss, while the FP method needs a perfectly polished end-face to avoid the impact of coupling losses on the extraction process. An alternative way to extract loss that can cancel out the role of input/output coupling efficiency is measuring the losses (extracted from the measured Q factor) of optical racetrack (RT) resonators with different dimensions. Waveguide losses can then be extrapolated from the propagation loss of the straight waveguide forming the RT [21]. In the following two sections, we will discuss the details on the fabrication and optical characterization of RT resonators and the extraction of waveguide losses in Y-cut LNOI substrates.

### *3.1 Fabrication*

The fabrication flow is shown in Fig. 3(a). We used three Y-cut LNOI samples with 1 μm, 1.5 μm and 2 μm thick BOX respectively, to explore the dependence of waveguide loss on BOX thickness. First, photonic patterns including grating couplers (GC), feeding waveguides ("U" shaped) and RT resonators, were patterned on CSAR 62 positive resist by electron-beam lithography. Among all the techniques to produce photonic devices on LNOI, physical etching using Ar plasma features high anisotropic etching profile. Therefore, we etched the LNOI film by inductively coupled plasma reactive ion etching (ICP-RIE) using the following parameters: Ar flow of 30 sccm, bias power of 100 W, ICP power of 600 W and pressure of 5 mT. The etching depth ($T_{etch}$) was about 300 nm. This recipe exhibited a selectivity to CSAR 62 resist of 1:1.

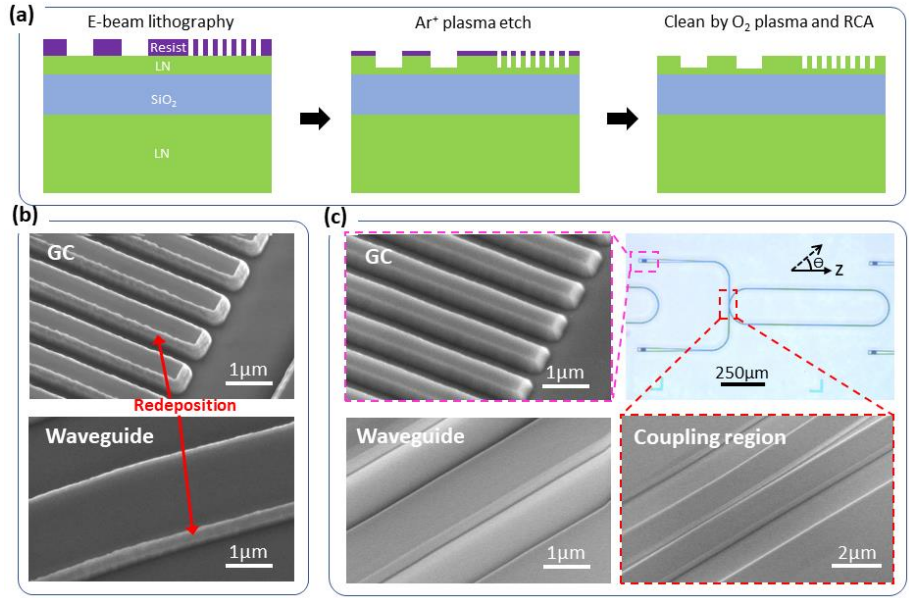

Fig. 3. (a) Fabrication flow of LNOI photonic devices. The LNOI wafer was purchased from NGK Insulators, LTD. (b) SEM pictures of GC and waveguide before RCA cleaning. The redeposition of LN along the sidewalls of the patterned features is clearly visible. (c) Microscope picture of one of the fabricated RT resonators (upper right) and SEM pictures of various photonic components after RCA cleaning. The waveguide is 2 μm wide and the coupling region (lower right) between the waveguides features a gap of 200 nm. The sidewall is very smooth after RCA cleaning. θ is the angle between the propagation direction of the straight waveguide and the crystalline z-axis (θ = 0° for the RT shown in the microscope picture).

The electron-beam resist left on the sample was removed by oxygen plasma cleaning right after etching. The final step included cleaning the etched patterns by RCA ($NH_4OH$, $H_2O_2$ and $H_2O$ mixed by the volume ratio of 1:1:5) at 60°C for 30 min. This RCA step is critical as it removes the organic residues and insoluble particles left behind by the etch step by changing their zeta potentials [26]. From the SEM images of the etching profile before and after RCA cleaning (Fig. 3(b) and (c)), it was obvious that the RCA cleaning effectively stripped the byproducts of the LN etch process and any residual resist that was stuck to the waveguide sidewalls.

*3.2 Characterization*

To characterize RT resonators, we coupled light in/out of the device by GCs shown in Fig. 3(c). The insertion loss of each GC was measured to be 7 dB at the wavelength of 1550 nm around which we characterized the RT resonators. By sweeping the wavelength of the tunable laser, the transmission spectra of RT resonators were recorded by a power-meter.

The RT resonators had fixed bend radius (R = 100 μm) but varying straight waveguide lengths (L = 300 μm, 700 μm, 2000 μm and 4000 μm. Total length of RT being $L_{tot}$ = $2\pi R + 2L$) and widths (W = 2 μm and 1 μm). First, we measured the transmission spectra of these RTs with different lengths and then extracted their intrinsic losses (loss in a single-pass trip besides coupling loss), $\alpha$, by fitting the experimental curves to a Lorentzian function. The transmission of RT resonators (L = 4000 μm) fabricated on a 2 μm thick BOX ($T_{BOX}$ = 2 μm) with W = 2 μm and 1 μm are shown in Fig. 4(a) and (b), respectively. The unloaded Q ($Q_U$) for these devices are higher than 1,000,000 for the case in which W = 2 μm. By comparing $\alpha$ from RTs with different L (or $L_{tot}$), we could eliminate the impact of bent waveguide loss and get the straight waveguide losses by simply looking at the slope of the fitted curve plotting loss vs. RT length (see Fig. 4(e) and (f)). The propagation loss of the straight waveguide with

W = 2 µm (1 µm) was as low as 0.37 ± 0.02 dB/cm (0.77 ± 0.07 dB/cm), which was among the lowest propagation loss reported so far for dry-etched LNOI waveguides with similar dimensions and without any over-cladding oxide layer [27]. The bending loss (100 µm radius) for W = 2 µm (1 µm) is extracted to be 1.1 ± 0.1 dB/cm (1.55 ± 0.3 dB/cm). Using an over-cladding layer could further reduce the scattering loss due to the lower index contrast between LN and the surroundings but requires additional steps (e.g. lithography and etching) to open windows to enable electrical contacts to external signal sources if such waveguides are to be used in practical applications for electro-optic (EO) or acousto-optic (AO) modulators.

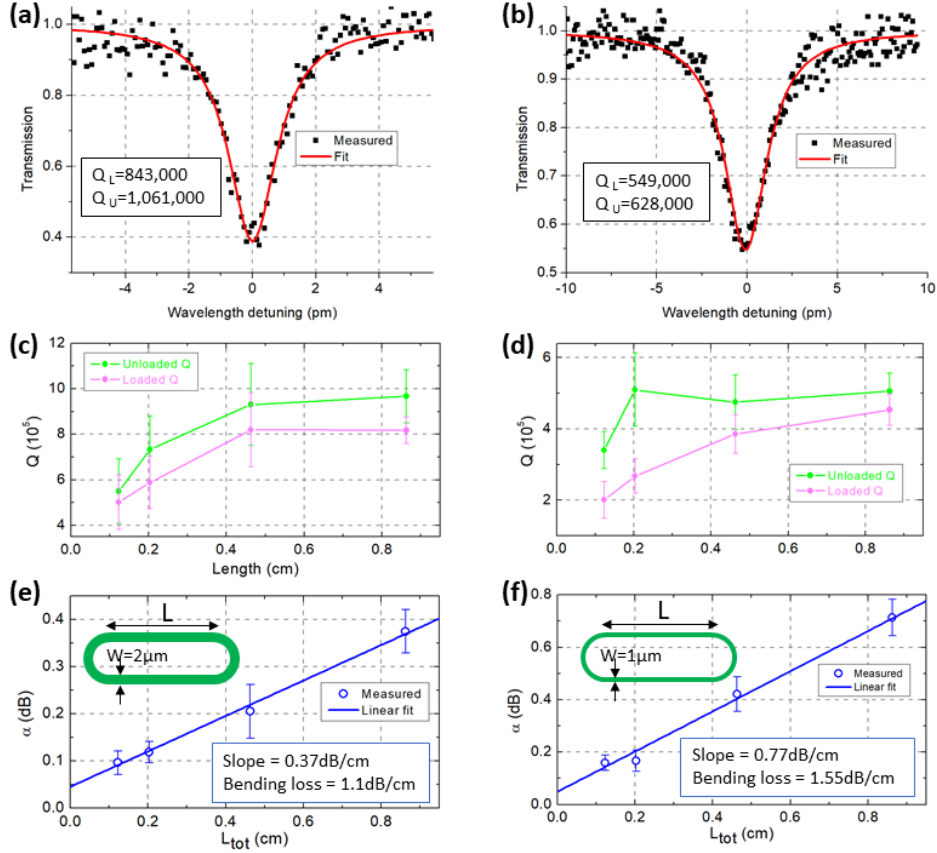

Fig. 4. (a) and (b): Measured (black dots) and fitted (red curves) transmission of RT resonator (L = 4000 µm) with W = 2 µm and W = 1 µm, respectively. (c) and (d): Loaded and unloaded Q factors in RT resonator as a function of the total length ($L_{tot} = 2\pi R + 2L$) for 2 µm and 1 µm wide straight waveguides respectively. (e) and (f): Intrinsic loss in RT resonator as a function of the total length. Average propagation losses of 0.37 dB/cm and 0.77 dB/cm for 2 µm and 1 µm wide straight waveguides were extracted by linearly fitting the experimental data. The variations in the value of the average propagation losses are set equal to one standard deviation of the losses extracted from several resonances around 1550 nm.

Using the same method reported above for extracting propagation losses, we measured the dependence of waveguide losses on different thicknesses of the BOX layer ($T_{BOX}$ = 1 µm, 1.5 µm). The measurement results are summarized in Table 1. It is clear that waveguide loss decreases with increasing $T_{BOX}$ independently of the width of the waveguide as predicted by simulations. Note that, for small $T_{BOX}$, *i.e.* $T_{BOX}$ = 1 µm, the leakage loss dominates over all other loss sources in the waveguides because the simulated value closely matches the experimental data. As $T_{BOX}$ increases, the energy leakage into the substrate becomes negligible and other loss factors like scattering loss become dominant. In particular, for the cases of $T_{BOX}$ = 2 μm and 1.5 μm, losses displayed in the table can be almost exclusively attributed to scattering losses since the leakage obtained from simulation is much less than the measured loss. This is a very important finding as it points out that $T_{BOX}$ of ~ 1.5 μm is sufficient to ensure low propagation loss without significantly impacting the electromechanical performance of SAW devices.

Since the design of high-performance EO or AO devices requires rotation of the waveguide with respect to the X-axis of the Y-cut crystal [17], we also measured the loss of the waveguide with $T_{BOX}$ = 2 μm for other two propagation directions (θ = 45° and 90°). We summarize these results in Table 1. For the case of W = 1 μm, it appears that the losses increase as θ increases from 0° to 90°. We speculate that since the value of $n_{eff}$ lowers as θ increases due to the birefringence of LN, then the smaller waveguides tend to suffer more from scatting losses as light experiences greater interaction with the waveguide sidewall.

Table 1. Summary of the waveguide propagation losses (Unit: dB/cm)

|  | $T_{BOX}$ = 1μm | $T_{BOX}$ = 1.5μm | $T_{BOX}$ = 2μm | | |
| --- | --- | --- | --- | --- | --- |
|  | θ = 0° | θ = 0° | θ = 0° | θ = 45° | θ = 90° |
| W = 1μm | 3 ± 0.5 | 0.94 ± 0.01 | 0.77 ± 0.07 | 0.83 ± 0.08 | 1.09 ± 0.07 |
| W = 2μm | 1.85 ± 0.4 | 0.59 ± 0.03 | 0.37 ± 0.02 | 0.23 ± 0.01 | 0.34 ± 0.02 |

## 4. Conclusions

In conclusion, we experimentally investigated the dependence of waveguide propagation loss on BOX thickness, waveguide width and in-plane orientation for Y-cut LNOI. As predicted by FDTD simulations, the measured loss decreased with increasing $T_{BOX}$ and reached values < 0.4 dB/cm when $T_{BOX}$ was selected to be 2 μm. By comparing the simulated and measured losses, we can conclude that leakage losses are dominant when $T_{BOX}$ is small, while, other loss sources (*i.e.* scattering loss) are more relevant when $T_{BOX}$ is large. Effectively, this work shows that for oxide thicknesses exceeding 1.5 μm, propagation losses due to substrate leakage can be dramatically reduced. Since, as evidenced by COMSOL simulations, the $k_t^2$ of SAW devices tends to deteriorate as $T_{BOX}$ increases, 1.5 μm of oxide could be considered as a good compromise for the making of high-performance AO devices. Ultimately, the selection of the oxide thickness will depend on the specific application and acoustic frequency for which the AO device is used. Nonetheless, this work reports important experimental insights on how propagation losses are affected by BOX thickness in Y-cut LNOI and offers general guidelines in the selection of the BOX thickness for high performance AO devices such as the inertial sensor reported in [13].

## Funding

This material is based upon work supported by the DARPA PRIGM-AIMS program under Award No. N66001-16-1-4025. Any opinions, findings, and conclusions or recommendations expressed in this publication are those of the authors and do not necessarily reflect the views of DARPA.

## Acknowledgments

Authors acknowledge S. Hiramatsu and Dr. Yuji Hori from NGK Insulators, LTD. for providing the LNOI wafers.